# Ferromagnetic Resonance in Permalloy Metasurfaces

N. Noginova[1*], V. Gubanov[2], M. Shahabuddin[1], Yu. Gubanova[2], S. Nesbit[1], V. V. Demidov[3], V.A. Atsarkin[3], E. N. Beginin[2], A. V. Sadovnikov[2]

[1]*Norfolk State University, Norfolk VA 23504 USA*

[2]*Laboratory "Metamaterials", Saratov State University, Saratov 410012, Russia*

[3]*Kotel'nikov Institute of Radio Engineering and Electronics, RAS, Moscow 125009, Russia*

[*]*corresponding author. email: noginova@nsu.edu*

**Abstract.** Permalloy films with one-dimensional (1D) profile modulation of submicron periodicity are fabricated based on commercially available DVD-R discs and studied using ferromagnetic resonance (FMR) method and micromagnetic numerical simulations. The main resonance position shows in-plane angular dependence which is strongly reminiscent of that in ferromagnetic films with uniaxial magnetic anisotropy. The main signal and additional low field lines are attributed to multiple standing spin wave resonances defined by the grating period. The results may present interest in magnetic metamaterials and magnonics applications.

**Introduction.**

Advances in nanofabrication and development of metamaterial concepts bring to life a new class of composite materials whose properties are artificially engineered via responses of nano-features to the electric component of electromagnetic field [1,2]. Nanoinclusions made from plasmonic metal (silver, gold, aluminum) are commonly used in optical metamaterials due to their strong response to illumination in the range of plasmon resonances [2]. Using magnetic inclusions which respond to the magnetic component of the electromagnetic radiation, one can design magnetic properties and propagation of electromagnetic waves in a material as well [3-8]. In similarity with optical metamaterials [2], such systems can be described in terms of the effective media approximation [9] with the material constants: dielectric permittivity, $\varepsilon$, and magnetic permeability, $\mu$. However, magnetic metamaterials are restricted to the radio frequency range since the dynamic of the magnetic response is relatively slow, with typical frequencies of magnetic resonance in GHz range. On the other hand, the use of magnetic components enables easy tunability of such systems with the external magnetic field. The effective permeability of composites with single-domain small magnetic nanoparticles in polymer matrices can be tuned from negative to positive values [5, 6], presenting interesting opportunities for temporal and spatial control of the wave propagation. Via mutual arrangement of magnetic nanofeatures and their shape, one can design materials with certain anisotropy of magnetic and microwave properties [10-15].

With increased sizes of magnetic features, the effective medium approximation using a single $\mu$ might be no longer sufficient as the response to the illumination becomes more complicated due to the excitation and propagation of spin waves. Structures with periodic arrangement of magnetic elements are commonly considered in terms of magnonic crystals [16-23] in analogy with photonic crystals [24, 25] exhibiting effective photonic forbidden and allowed bands. In this work we study grating-like permalloy structures with the submicron periodicity. Permalloy is a soft ferromagnetic with high magnetic permeability; it is a common material for various magnetic and magnonic [13,22] studies. In addition, it exhibits plasmonic behavior [26] and an interesting coupling between plasmonic, magnetic and electric properties [27], presenting interest for plasmon-induced magnetization switching and magnetically controlled plasmonics. In our work, we employ the FMR method and micromagnetic numerical simulations in order to better understand magnetic and magnetic resonance behavior of such systems, applicability of metamaterial description, and a role of spin wave-related effects.

## Experimental

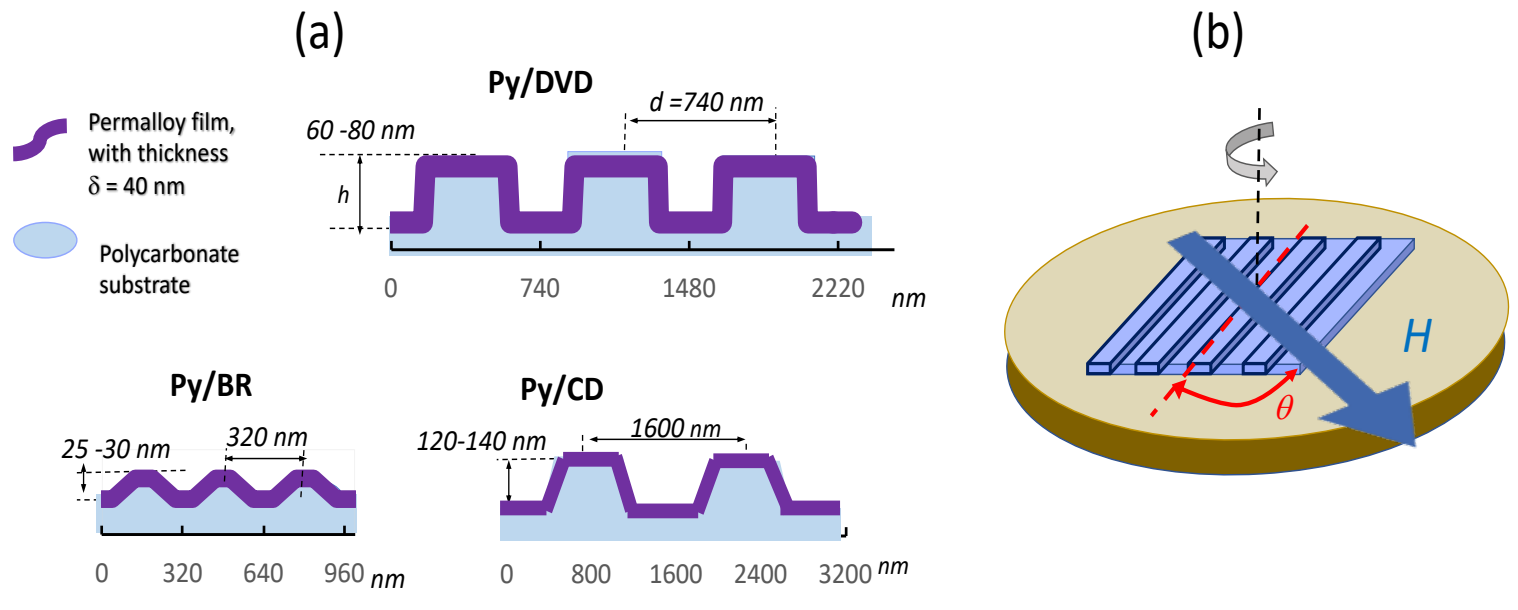

Fig. 1. (a) Profile schematics of Py/DVD, Py/BR and Py/CD structures (as indicated). (b) Orientation of the sample in FMR experiments.

Our experimental structures are permalloy (Ni-Fe alloy with 80% of Ni and 20% of Fe) thin films with one-dimensional (1D) profile modulation based on the substrates derived from commercially available DVD-R discs. The fabrication starts with obtaining polycarbonate grating substrates from disassembling commercial DVD-R by carefully taking out the polymer, plastic, silver and protective coating layers. Then, permalloy (Py) with a thickness $\delta = 40$ nm is deposited on the prepared and precut DVD substrates using e-beam evaporation. The thickness of the film is independently tested with a profilometer by measuring films simultaneously deposited on glass substrates. Atomic force microscopy (AFM) confirms the profile modulation parameters, the periodicity $d$ = 740 nm and modulation height $h$ = 60-80 nm. As main experiments and numerical simulations are performed with Py/DVD structures, for comparison purposes we also prepared permalloy structures with different profile-modulation parameters using substrates derived from Blu-ray or CD discs, with $d$ = 320 nm and $h$ = 25-30 nm in Py/BR structures, and $d$ = 1600 nm and $h$ = 120-140 nm (Py/CD), see Fig. 1 (a) for the schematics.

Ferromagnetic resonance curves are recorded using the Bruker EPR Spectrometer at 10 GHz microwave frequency (X Band). The sample is placed inside the microwave cavity with the sample orientation corresponding to the external magnetic field, $H$ in plane of the film, Fig 1(b). The direction of the grooves makes an angle $\theta$ with $H$. The signal (the derivative of the microwave absorption vs $H$) is recorded for various $\theta$.

Typical ferromagnetic resonance spectra observed in Py/DVD are shown in Fig.2 (a). (In this plot, the field is shown in Oersted as is in the original recordings.) The signal position depends on the orientation angle, shifting from the lowest to highest field when $\theta$ changes from 0 to 90 deg, corresponding, respectively, to the parallel and perpendicular orientations of the grooves in respect to the external field.

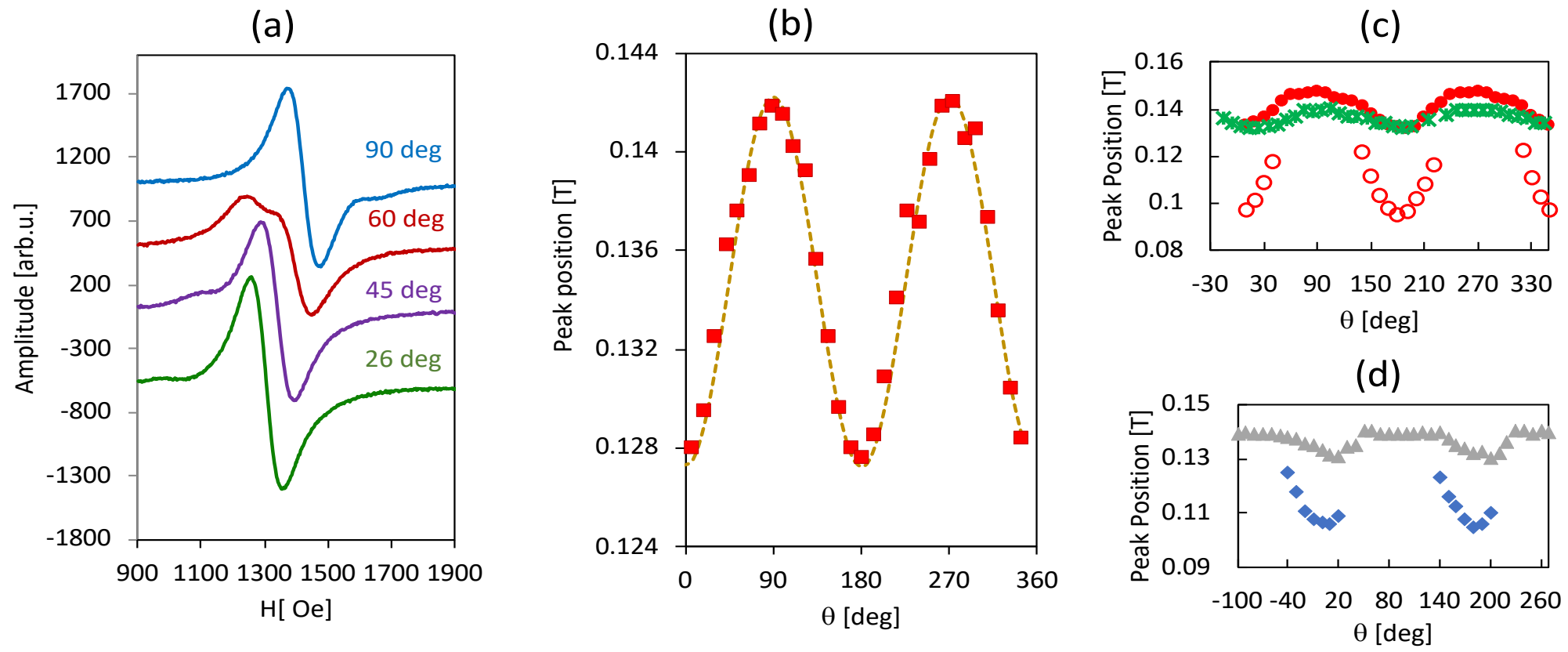

Fig. 2. (a) Typical FMR spectra and (b) position of the main peak as the function of $\theta$ in Py/DVD. Dashed trace is fitting with Eq. 1. (c) Peak positions in Py/CD (stars) and Py/DVD (circles), main peak (closed symbols), additional component (open symbols). (d) Positions of two peaks in Py/BR vs angle.

The position of the main FMR peak (determined from fitting with the Lorentzian lines) is plotted as the function of $\theta$, Fig. 2 (b). The dependence closely follows $cos\ 2\theta$ function. An additional component can be clearly distinguished in some orientations, with the position strongly dependent on the angle as well. The relative strength of this component slightly varies from sample to sample while the angular behavior (Fig. 2 (c), open circles) follows the same cos $2\theta$ dependence.

We explored the FMR in the other structures as well. A single peak in the Py/CD structures demonstrates the same angular dependence (Fig. 2(c), stars) but with a smaller amplitude. The behavior of Py/BR is different. In a broad range of angles ($\theta > 40^o$), a single FMR peak is observed with practically no dependence on the orientation. Splitting into two peaks and a strong angular dependence is observed for low angles. Note, that in the current work we concentrate on the Py/DVD system; the other systems will be studied in detail elsewhere.

**Discussion and Modeling**

The angular dependence of the main FMR signal in Py/DVD is strongly reminiscent of that in ferromagnetic films with the growth-induced in-plane uniaxial anisotropy [28,29]. This can be expected, taking into account that the structural geometry of our gratings is comparable with that of crystalline films [28] but with a submicron-size of the features instead of interatomic distances. Note that films with this type of the magnetic anisotropy exhibit a sharp reorientation of the magnetization upon a small increment of the magnetic field [30, 31], and are of particular interest for optically induced magnetization switching via angular momentum transfer from light to matter [32, 33].

Assuming a relatively small magnetic anisotropy $H_p < H < M$, the FMR condition in our excitation geometry reads [34],

$$\left(\frac{\omega}{\mu_0\gamma}\right)^2 = \left(H + M + H_p \cos^2\theta\right)\left(H + H_p \cos 2\theta\right), \quad (1).$$

predicting the periodic dependence with the period of 180°, minimum at 0° and maximum at 90° as has been observed in the experiment. Here $\omega$ is the FMR frequency, $\gamma$ is the gyromagnetic ratio, $M$ is the magnetization, and $H_p$ is the anisotropy field [35]. Fitting the experimental data with Eq. 1, the effective anisotropy fields are estimated as $\mu_0 H_p$ = 7.6 mT in the Py/DVD and 3.7 mT in Py/CD, assuming the saturation magnetization of permalloy, $M = M_S = 6.4 * 10^5$ A/m. This is in agreement with the literature [12-15] as well, where patterned magnetic structures such as films deposited on the grating-like substrates [13] exhibit a uniaxial anisotropy with the easy axis parallel to the direction of grooves. However, in our Py/DVD systems, additional features are clearly seen, which could not be described with this simple approximation.

In order to better understand magnetic behavior of the Py/DVD structure, we perform numerical simulations, considering a meander-like structure with the following parameters: saturation magnetization of permalloy $M = M_S = 6*10^5$ A/m, periodicity $d = 740$ nm, modulation height $h$ = 80 nm, thickness of permalloy, $\delta$ = 50 nm at horizontal stages, and various thicknesses $w \leq \delta$ at vertical walls. Since the deposition of metal on the top of the substrate can produce vertical walls with reduced thickness, additional simulations have been performed to explore the role of reduced $w$. As we found, small variations in the range of possible thicknesses $0 < w \leq \delta$ do not significantly affect the results. The detailed study of FMR behavior in patterned systems with different parameters including variations in the periodicity, film thickness and the shape of the profile modulation (sine-wave or rectangular) is the subject of a separate study. In the numerical simulations, w e apply an approach discussed in detail in [19] and perform micromagnetic modeling of our structure using the MuMax3 software [36]. The calculation is based on the Landau-Lifshits-Gilbert equation with damping parameter of 0.007. We consider the same geometry as in the experiment: the external magnetic field lies in plane with the structure and the angle, $\theta$, between the direction of grooves and field varies from $0^0$ to 90. The microwave field with the magnitude of $10^{-4}$ T is applied along $y$ axis (perpendicular to the **H** direction).

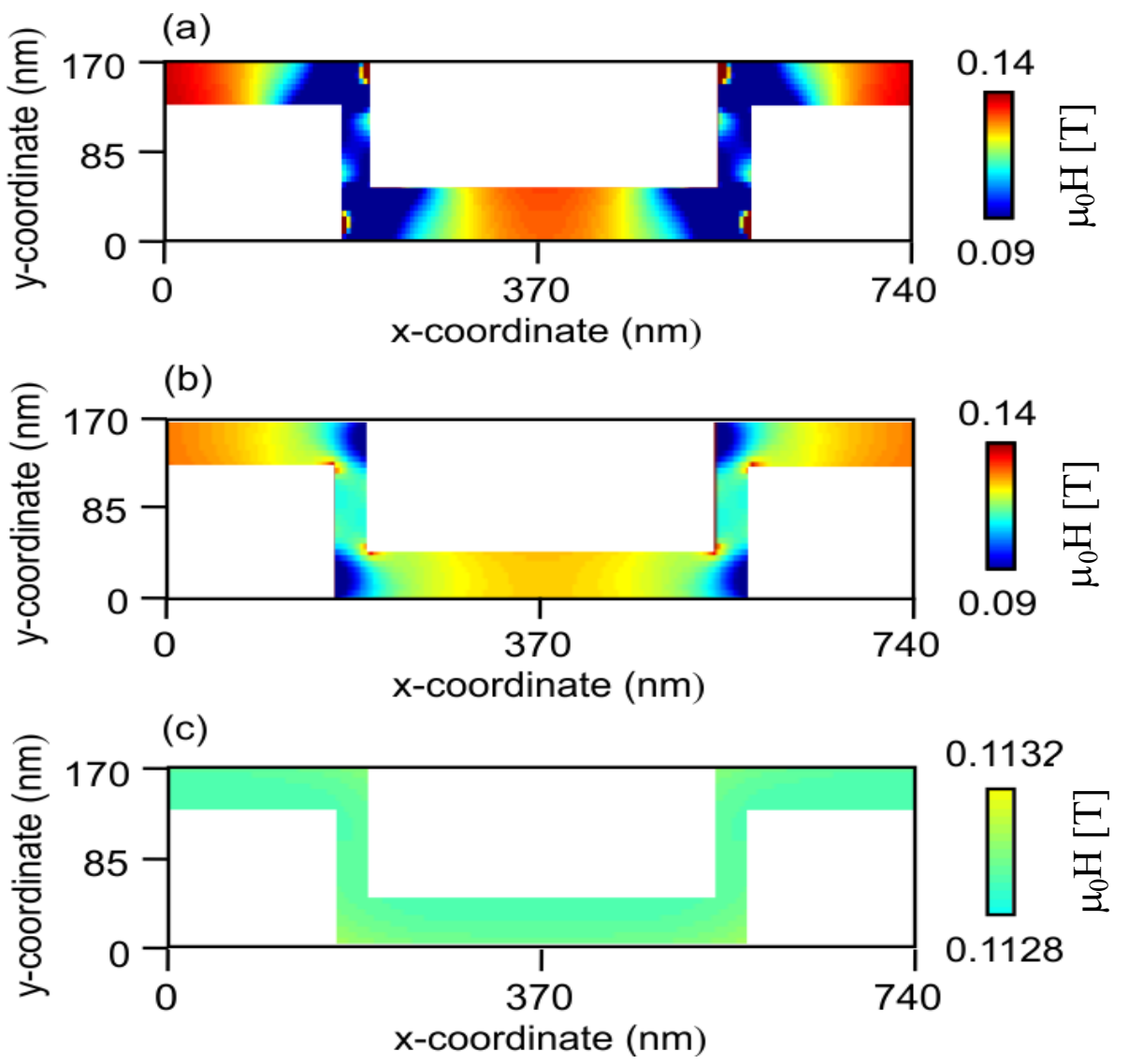

Fig. 3. Distribution of the internal magnetic field at different rotation angles: (a) $\theta$ = 90°, (b) $\theta$ = 45° (c) $\theta$ = $0^0$. $\mu_0 H_0 = 0.2$ T.

First we estimate local distributions of the static magnetization $M$ and effective internal fields $H_i$ (resulting from external and demagnetization fields). In Figure 3, distributions of $H_i$ (absolute values) are shown at various rotation angles: when the external magnetic field is perpendicular to the direction of grooves (Fig. 3(a)); makes an angle of 45° (Fig. 3 (b)) or parallel to the grooves (Fig, 3(c)). As one can see, at $\theta = 90^0$ and 45°, the magnetic response of the material is strongly modulated in space with maxima of $H_i$ observed in the middle of horizontal stages and minima observed at the corners. The magnitude of this spatial modulation decreases with the decrease in the angle, vanishing at the parallel orientation of the grooves and external field, Fig. 3 (c). Absorption at the 10 GHz frequency is calculated following the approach [19] as the function of the external magnetic field. In a flat film, a single resonance peak is expected at $\mu_0 H_0$ = 0.142 T, while in the profile- modulated structures, several peaks can be resolved with the positions dependent on the orientation.

Three major peaks predicted in the Py/DVD structure (with the period $d$= 740 nm and wall thickness $w$= 12.5 nm) are shown in Fig. 4 (a). With an increase in $\theta$, these peaks shift toward $H_0$, Fig. 4 (b). The angular dependences for the Peak 1 and Peak 2 positions fairly well correspond to those of the main and additional peaks observed in experiment, Fig. 4(c).

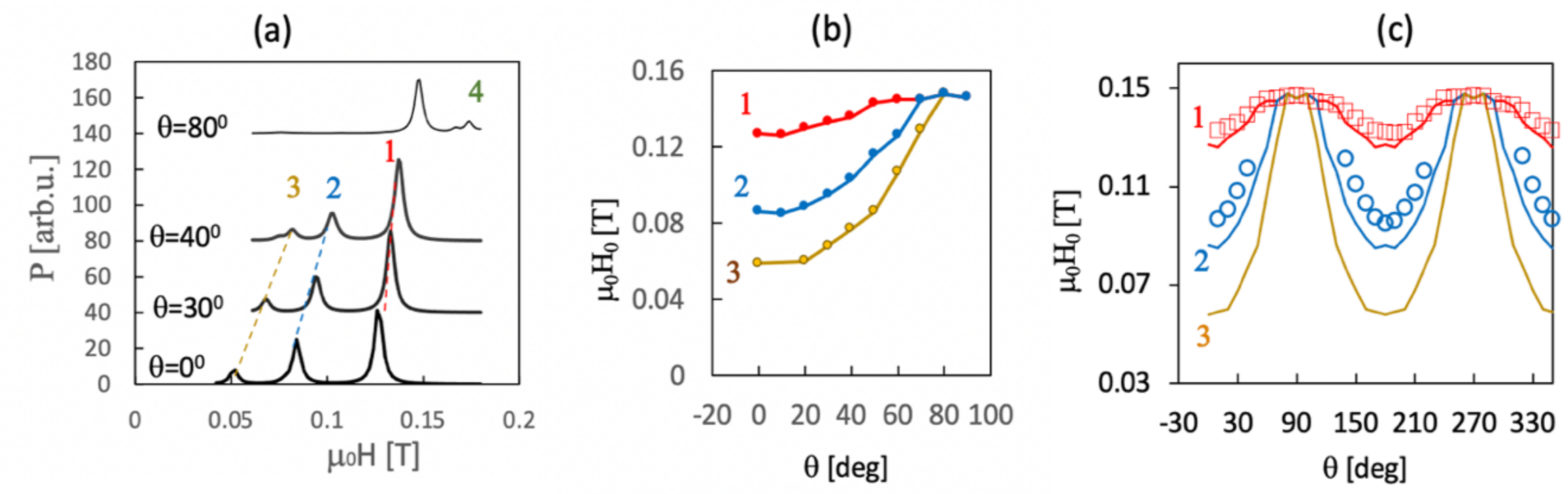

Fig. 4. (a) Simulated microwave (10 GHz) absorption vs field in Py/DVD . (b) Peak positions vs orientation angle, (c) Comparison with experiment. Experimental data are shown with symbols, results of numerical simulations are solid traces. Numbers 1, 2, 3 indicate corresponding peaks.

Let us assume that the peaks observed below $H_0$ both in simulations and experiments correspond to the resonance modes formed by the magnetic surface spin waves (MSSW [17-19, 35]). (The Peak 4 in Fig. 4 (a) with the position above $H_0$ is likely related to the volume modes and is not discussed here). When the field **H** and the magnetization, **M**, are directed along the grooves ($\theta$ = 0), MSSW waves are excited in the perpendicular direction, see Fig.5(a) which illustrates the amplitude of the precessing transverse magnetization $m(t)$ under the resonance conditions. Since the spatial modulation (defined by the grating period) has the period of $d$ in this direction, the resonance condition is expected at

$$k = \pi N/d, \qquad (2)$$

where $N$ = 1,2,3.. is an integer. If the magnetization is directed under an angle $\theta$ in respect to the groove direction (Fig 5 (b)), the period of the modulation becomes $d/\cos\theta$, and the k vector of the resonance mode can be found from the condition,

$$k = N\frac{\pi}{d}\cos\theta. \qquad (3)$$

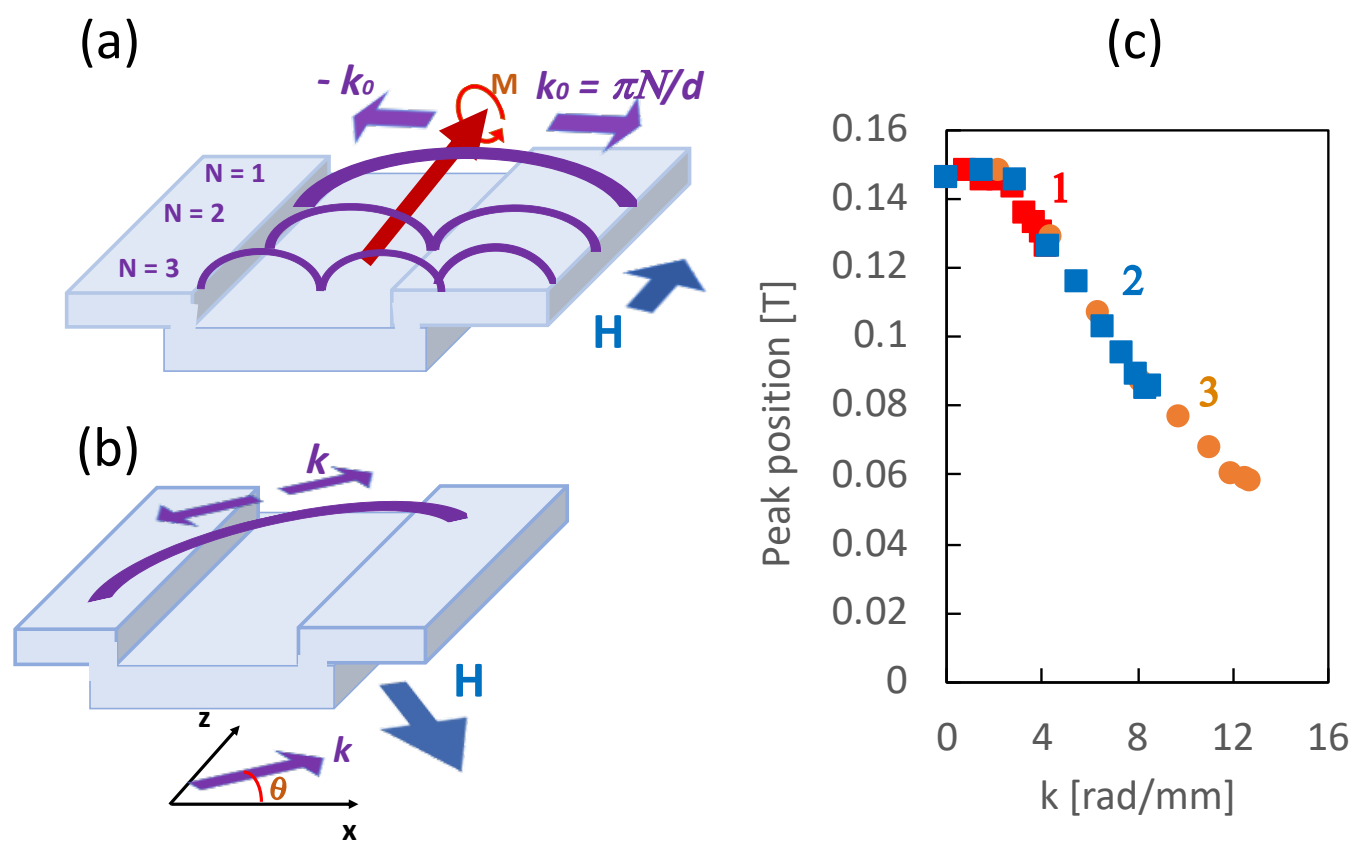

Fig, 5. (a, b) Standing modes at (a) perpendicular orientation and (b) an arbitrary angle; (c) Resonance field vs *k*, simulations. The numbers 1, 2, 3 indicate corresponding datasets of Fig. 4.

Let us replot the peak position estimated from the numerical calculations (Fig. 4 (b)) as the function of the k-vector estimated from the Eq. 3 with $N = 1$ for the dataset 1, and $N = 2$ and $N = 3$ for the datasets 2 and 3 correspondingly. All the data fits the single curve (Fig. 5 (c)) confirming our assumptions above. Note, that, alternatively, similar speculations can be applied assuming a twice smaller period *d/2* of magnetic modulation (which corresponds to the width of each horizontal segment). In this case, the graph (c) will be stretched twice along the *x* axis. As discussed below, our data rather indicate that *d* is the period of modulation.

In frames of the same numerical approach, instantaneous values of the dynamic (transverse) magnetization component $m_x$ are calculated under the resonance conditions of Peaks 1, 2 and 3 at $\theta = 0$, see the panels on the left sides of Fig. 6 (a-c) and their analysis on the right. These patterns are rather complicated, in particular, inside the vertical walls. However, the presence of standing waves is evident, with the *k* vectors of $\frac{\pi}{d}$, $2\frac{\pi}{d}$ and $3\frac{\pi}{d}$ corresponding to the resonances 1, 2 and 3 respectively.

We put together theory and experiment in Figure 6 (d). Points are the experimental results with the abscissas calculated as following: Red circles and triangles: Py/DVD ($d_{DVD}$ = 740 nm). The k vectors are calculated as $k = \frac{\pi}{d_{dvd}} N \cos\theta$, $N = 1$ for the main peak (circles), and $N = 2$ for the second peak (triangles). Green stars: Py/CD ($d_{CD}$ = 1600 nm). The k vectors are calculated at the same manner assuming $N = 1$, $k = \frac{\pi}{d_{cd}} \cos\theta$. We make an attempt to add the results obtained in Py/BR ( $d_{BR}$ = 320 nm) as well. Points (blue squares) fit the general tendency if we use the second (smaller) peak (which shows strong angular dependence) and estimate $k = \frac{\pi}{d_{BR}} \cos\theta$. The dashed curve is obtained from the numerical simulations ( figure 5(c)). The solid curve is shown for comparison; it is obtained from Eq. 4 which describes the dispersion curve of the MSSW for the flat film [37] assuming the film thickness $\delta$ = 50 nm,

$$\left(\frac{\omega}{\mu_0 \gamma}\right)^2 = H(H+M) + \frac{M^2}{4}\left(1 - e^{-2k\delta}\right). \qquad (4)$$

In principle, assuming a certain relationship between the field and the resonance frequency one can derive the dispersion curve of the spin waves in our structures from the data plotted in Fig. 6. It is already seen that at higher $k$, the dispersion curve is close to that in a flat film. This is expected since the spin waves with smaller wavelengths are less affected by the profile geometry.

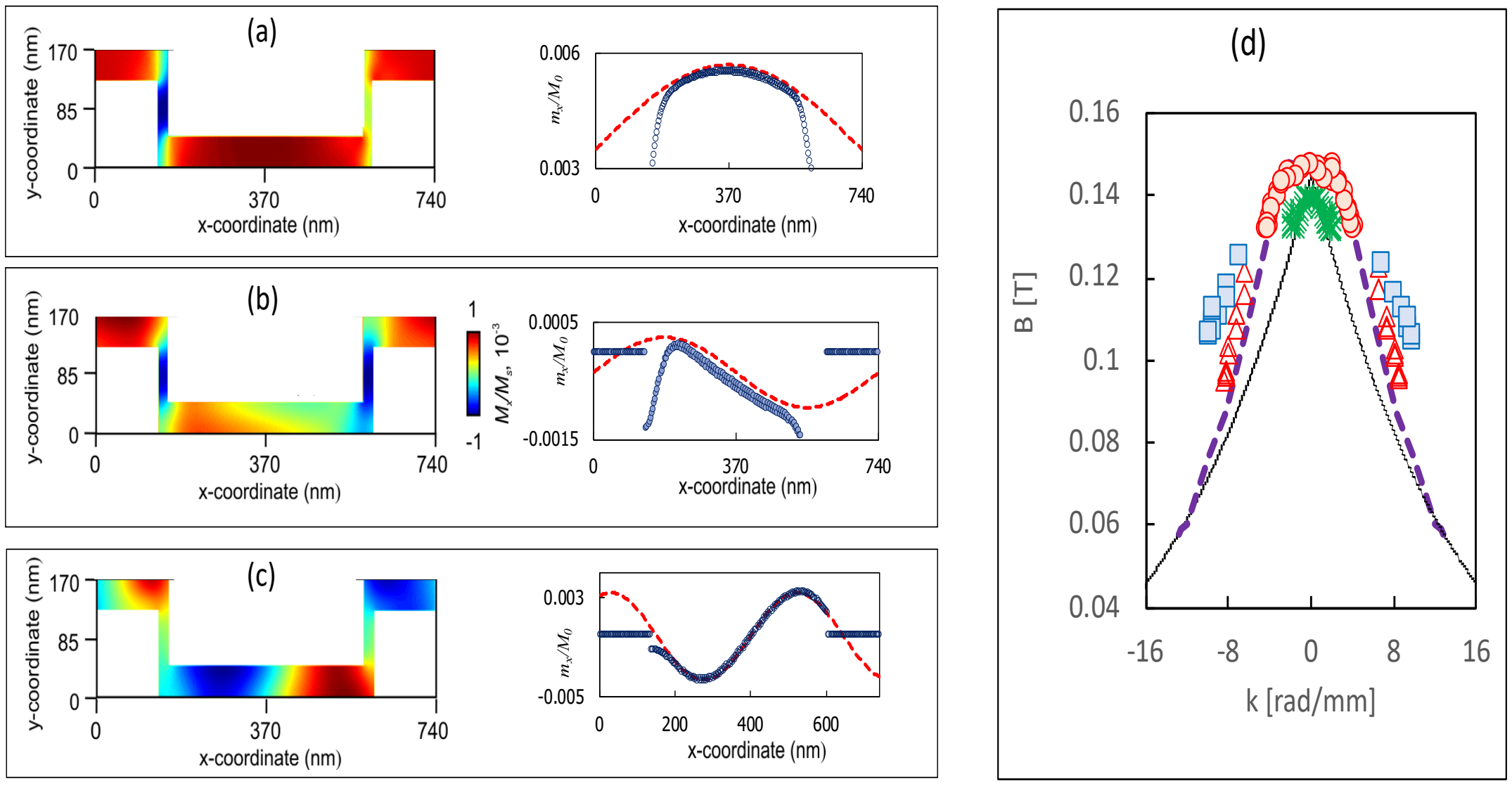

Fig. 6. (a-c) Distributions of dynamic magnetization $m_x$ (left panels) and profiles of $m_x$ along the half-depth of the lower horizontal segment (plots on the right) at the resonance conditions of Peaks 1, 2 and 3 at $\theta = 0$, (a) $\mu_0 H$ =0.126T, (b) $\mu_0 H$ = 0.0835 Tm and (c) $\mu_0 H$ =0.052T. Red dashed traces are sine waves with the periodicity of (a) 2$d$, (b) $d$, and (c) 2$d$/3.
(d) Experiment (points) and theory (traces). Py/DVD, main peak (circles), additional peak b: (triangles); Py/CD (stars), Py/BR (squares). Numerical simulations (dashed trace) and predictions for a flat film, Eq (4) (solid trace).

In conclusion, 1D profile-modulated permalloy films (continuous gratings) are studied with the ferromagnetic resonance method. The position of the main resonance is found to be strongly dependent on the angle between the direction of the grooves and the external magnetic field (which is kept in-plane). The angular dependence is very similar to that observed in crystalline films with uniaxial magnetic anisotropy and in-plane easy axis. Additional resonances of a smaller magnitude are observed at lower fields, which show a similar angular dependence as well. The findings are discussed in terms of MSSW resonances and confirmed by micromagnetic simulations.

## Acknowledgements

N. Noginova, M. Shahabuddin and S. Nesbit would like to acknowledge financial support from National Science Foundation (NSF) (1830886), Air Force Office of Scientific Research(AFOSR) (FA9550-18-1-0417) and Department of Defence (DoD) (W911NF1810472). The work of authors from Russian Federation is carried out within the framework of the state task and partially was supported by Russian Foundation for Basic Research, projects No. 18-57-16001.